\documentclass[english,12pt]{article}
\usepackage{array}
\usepackage{graphicx}
\usepackage{amssymb}
\usepackage{amsmath}
\usepackage{multirow}
\usepackage{prettyref}
\usepackage{babel}
\usepackage{units}
\usepackage[latin1]{inputenc}
\usepackage{amsfonts}
\usepackage{amssymb}
\usepackage{babel}
\usepackage{perpage} 
\MakePerPage{footnote} 
\usepackage{cite}
\def\@fmsl@sh#1#2#3{\m@th\ooalign{$\hfil#1\mkern#2/\hfil$\crcr$#1#3$}}
 \def\eq#1\en{\begin{equation}#1\end{equation}}
\def\s[#1,#2]{[#1\stackrel{\star}{,}#2]}
\def\sx[#1,#2]{[#1\stackrel{\star_{x}}{,}#2]}

\newcommand{\nc}{\newcommand}
\nc{\beq}{\begin{equation}}
\nc{\eeq}{\end{equation}}
\nc{\beqa}{\begin{eqnarray}}
\nc{\eeqa}{\end{eqnarray}}

\def\bc{\begin{center}}
\def\ec{\end{center}}

\def\to{\rightarrow}

\def\gsim{\mathrel{\mathpalette\atversim>}}

\def\bc{\begin{center}}
\def\ec{\end{center}}

\def\gsim{\mathrel{\rlap{\lower4pt\hbox{\hskip1pt$\sim$}}

    \raise1pt\hbox{$>$}}}       

\def\gsim{\mathrel{\rlap{\lower4pt\hbox{\hskip1pt$\sim$}}
    \raise1pt\hbox{$>$}}}       

\begin{document}
\makeatletter
\def\fmslash{\@ifnextchar[{\fmsl@sh}{\fmsl@sh[0mu]}}
\def\fmsl@sh[#1]#2{%
  \mathchoice
    {\@fmsl@sh\displaystyle{#1}{#2}}%
    {\@fmsl@sh\textstyle{#1}{#2}}%
    {\@fmsl@sh\scriptstyle{#1}{#2}}%
    {\@fmsl@sh\scriptscriptstyle{#1}{#2}}}
\def\@fmsl@sh#1#2#3{\m@th\ooalign{$\hfil#1\mkern#2/\hfil$\crcr$#1#3$}}
\makeatother

\thispagestyle{empty}
\begin{titlepage}
\boldmath
\begin{center}
  \Large {\bf Quantum Mechanics,  Gravity and Modified Quantization Relations}
    \end{center}
\unboldmath
\vspace{0.2cm}
\begin{center}
{  {\large Xavier Calmet}\footnote{x.calmet@sussex.ac.uk}}
 \end{center}
\begin{center}
{\sl Physics $\&$ Astronomy, 
University of Sussex,   Falmer, Brighton, BN1 9QH, United Kingdom 
}
\end{center}
\vspace{5cm}
\begin{abstract}
\noindent
In this paper we investigate a possible energy scale dependence of the quantization rules and in particular, from a phenomenological point of view, an energy scale dependence of an effective $\hbar$. We set a bound on the deviation from the value of $\hbar$ at the muon scale from its usual value using measurements of the anomalous magnetic moment of the muon.  Assuming that inflation took place, we can conclude that nature is described by a quantum theory at least up to an energy scale of about $10^{16}$ GeV.
 \end{abstract}  

\end{titlepage}


\newpage

\section{Introduction}
There are three constants of nature which are usually assumed to be fundamental and  immutable. These are the speed of light $c$ Planck's constant  $h$ and Newton's constant $G_N$. The speed of light $c=299 792 458$ m/s \cite{PDG}, determines the velocity of a massless particle in vacuum and is introduced in special relativity as a conversion factor between time and space. It allows to define an invariant length element $ds^2= c^2 dt^2- dx_idx^i$. Planck's constant $h=6.626 069 57(29) \times 10^{-34}$ J$\cdot$s \cite{PDG} is the physical constant which enters quantum mechanics and its extension quantum field theory. Planck's constant allows to convert the energy $E$ of  a particle into the frequency $\nu$ of the wave associated with this particle using the relation $E=h \nu$. The momentum of the particle $p$ is related to de Broglie's wavelength $\lambda$ via $E= h c/\lambda$ where $c$ is the speed of light in vacuum. It is common to introduce Planck's reduced constant, $\hbar =h/(2\pi)=1.054571726(47)\times 10^{-34}$ J$\cdot$s \cite{PDG}. Note that there are different ways to measure this fundamental constant, see e.g. \cite{steiner}.
Among these constants, Newton's constant $G_N=6.67 \times 10^{-11}$ N$\cdot$(m/kg)$^2$ \cite{PDG} is on a slightly different footing since it fixes the strength of the gravitational interactions and it gets renormalized (see e.g. \cite{Larsen:1995ax}). It is thus dependent of the energy scale at which it is probed.

Planck's constant plays an essential role in the quantum mechanics and quantum field theory.
 It is this constant of nature which determines the amount of quantization both for the first quantization and second quantization. The quantization rules $[x_i,p_j]=i \hbar \delta_{ij}$ in quantum mechanics and e.g.  $[\phi({\bf x}),\Pi({\bf y})]=i \hbar \delta({\bf x} - {\bf y})$ in quantum field theory both involve $\hbar$.

In this paper, we will show that both the first and second quantization rules which consist in imposing non-trivial commutation relations for coordinates and momenta (in the case of the first quantization, i.e. quantum mechanics) and fields and their canonical momenta (in the case of the second quantization, i.e. quantum field theory) can be modified by quantum gravity or modification of quantum mechanics itself.

For example, in some models such as string theory, the uncertainty relation of quantum mechanics gets modified. Amati, Ciafaloni and Veneziano have shown \cite{Amati:1988tn,Amati:1992zb,Amati:1993tb,Amati:2007ak} by studying the scattering of massless strings in the trans-Planckian regime that the usual uncertainty principle of quantum mechanics 
\begin{eqnarray}
\Delta x \Delta p > \hbar
\end{eqnarray}
generalizes to  
 \begin{eqnarray}
 \Delta x \Delta p > \hbar +\alpha^\prime \Delta p^2,
 \end{eqnarray}
  where $\alpha^\prime=G_N/g^2$, $G_N$ is Newton's constant and $g$ is the string loop expansion parameter. This relation is called the generalized uncertainty principle.

  While this specific relation was derived in string theory similar ones appear in different models of quantum gravity, see e.g. \cite{Hossenfelder:2012jw} for a recent review. One expects such relation in most theories attempting to unify quantum mechanics and general relativity \cite{Calmet}. Indeed, black holes are expected to form in the collisions of particles at very high energy. For concreteness, let us consider the collision of two particles, it can be shown that even for a non-zero impact parameter a black hole will form at center of mass energies above the Planck scale \cite{CCW}. Imagine that we collide these particles at energies just above the threshold for black hole formation, i.e. just above the Planck mass. The black hole created will have a mass of the order of the Planck mass and thus a size (Schwarzschild radius) of the order of the inverse of the Planck mass. Such a scattering experiment is thus not able to resolve distances shorter than the Schwarzschild radius corresponding to the black hole with a minimal length of the order of the Planck scale.
 As we increase the center of mass energy, so does the mass of the black hole. The black hole becomes larger and the length that can be probed by the scattering experiment increases with energy as well. Thus, as in the case of ACV,  increasing the center of mass energy of the scattering experiment does not allow to resolve shorter distances as the $\Delta x$ probed by the scattering experiment increases with the center of mass energy. We thus expect generically a modification of the uncertainty relation of the type:
\begin{eqnarray}
\Delta x \Delta p > \hbar +\alpha f(\Delta p^2),
\end{eqnarray}
where the parameter $\alpha$ is positive. That is if quantum mechanics remains valid up to all energy scales.

Generically speaking one should be able to derive such generalized uncertainty relations from a modification of the commutation relation between $x$ and $p$ \cite{Hossenfelder:2012jw}:
\begin{eqnarray} \label{ncrel}
[x^\nu,p_\mu]=i \hbar \frac{\partial f(k)^\nu}{\partial k^\mu }
\end{eqnarray}
where $f(k)_\nu=p_\nu$ where $f_\nu$ are invertible functions such that their inverse $f^{-1}_\nu(p)=k_\nu$ are well defined.  
One way to think of such models is as of an energy scale $\mu$ dependent effective $\hbar$:
\begin{eqnarray}
[x^\nu,p_\mu]=i \hbar(\mu) \delta^\nu_\mu
\end{eqnarray}
in analogy to renormalized coupling constants such as the fine-structure constant of quantum electrodynamics. Note that the noncommutation relations in Eq. (\ref{ncrel}) should, as we shall see, be derived from a  modified Lie algebra. In this paper we shall investigate whether $\hbar$ is, as $G_N$,  dependent of the energy scale at which it is measured. Such an energy dependence would not be motivated by the necessity to renormalize the theory, but rather by a possible modification of quantum mechanics at short distance or equivalently at very high energy. 

Independently of our theoretical arguments, it is an interesting empirical question to investigate whether Planck's constant has an energy dependence. We have seen that most unifications of general relativity and quantum mechanics are likely to lead to an effective Planck constant that increases with energy. There is however another framework where the opposite is expected. This framework is that advocated by 't Hooft who has been investigating the possibility that quantum mechanics could be an emergent phenomenon \cite{'tHooft:2010zz,Hooft:2014kka}. While this proposal faces potential difficulties in explaining the Bell inequalities, it is worth looking at it from a phenomenological point of view. If quantum mechanics is an emergent phenomenon, in the sense that our world appears quantum mechanical while the underlying theory at short distance would be classical, $\hbar$ must effectively depend on the energy at which we probe it.
On quantum scale $\hbar$ takes its usual value while at the fundamental level $\hbar$ tends to zero and one recovers a form of deterministic physics, maybe in the form of cellular automata as envisaged by 't Hooft. It is worth noting that the limit $\hbar \to 0$ does not imply that loop diagrams in quantum field theory necessarily all go to zero. It has been pointed out that some parts of loop diagrams correspond to purely classical physics \cite{Holstein:2004dn,Brodsky:2010zk}. The Feynman diagram expansion, which leads to loop diagrams is an expansion in the coupling constant of the theory and not in $\hbar$. One should therefore be careful when considering this limit.

We will first show that the assumption that Lorentz invariance is the correct symmetry of nature leads to surprisingly little freedom in terms of how we can modify the value of  $\hbar$. We will then show that it is universal in a sense to be defined below.

\section{Modification of the quantization rules: universality of $\hbar$}

It is important to realize that the quantization rules cannot be modified in a arbitrary fashion. As we shall explain shortly, the quantization rules are not axioms of the theory but rather derived from the foundations of quantum mechanics and quantum field theory.  In particular we shall show that $\hbar$ must be universal. The universality of $\hbar$ is widely accepted, here we provide a proof. 

It is very well understood that the speed of light is universal, its value is the same in all laws of physics which are Lorentz invariant. Similarly Newton's constant is universal because of the equivalence principle. 
The reduced Planck constant $\hbar=h/(2\pi)$ appears both in quantum mechanics and in quantum field theory.  In quantum mechanics, one could naively have a different $\hbar$ in each of the quantization rules:
\begin{eqnarray}
[x,p_x]&=&i \hbar_x, \\  \nonumber
[y,p_y]&=&i \hbar_y,\\  \nonumber
[z,p_z]&=&i \hbar_z.
\end{eqnarray}
Requiring that $x$ and $p$ do not commute is sometimes called the first quantization. Note that as usual $[x_\mu,x_\nu]=0$ and $[p_\mu,p_\nu]=0$. The second quantization rule is relevant to quantum field theory, where for e.g. a real scalar field one imposes
\begin{eqnarray}
[\phi({\bf x}),\Pi({\bf y})]=i \hbar_2 \delta({\bf x} - {\bf y}),
\end{eqnarray}
where $\Pi(x) = \dot \phi(x)$ where the dot stands for a time derivative. Furthermore, one has $[\phi(x),\phi(y)]=0$ and $[\Pi_\mu(x),\Pi_\nu(y)]=0$. Why should the $\hbar$ in these expressions be identical and in quantum field theory, why should $\hbar$ be universal for all particle species? In other words, why should $\hbar$ be a universal constant of nature? It is important to remember that the quantization rules whether one is looking at the first or second quantization procedure are not axioms of the quantum mechanics or quantum field theory but rather are derived from the axioms of these theories.  Note that while we will work in the Schr\" odiger picture, the same reasoning obviously applies to the Heisenberg picture. We will show that there is no freedom at all and that because ultimately of the structure of the Lie algebra for the generators of the Lorentz group, $\hbar$ must be universal.

Let us first consider quantum mechanics.  We will use the derivation of the commutator relation presented in Weinberg's book \cite{Weinberg}.
We will show that translation invariance implies that the $\hbar$ is the momentum operator $P$ (i.e. in the Schr\" odinger equation) is the same as that of the commutator relation between the operators $X$ and $P$.

As usual, one assumes the existence an operator $\vec X_n$ corresponding to the observable $\vec x_n$, i.e. the position of a particle where $n$ labels the individual particles. We shall consider translation invariance (physics should not depend on a shift of e.g. the origin of the coordinate system). In other words, transformations $\vec X_n \to \vec X_n +\vec a$ where $\vec a$ is an arbitrary three vector should not modify the physics. Translation invariance is generate by a unitary operator $U(\vec a)$:
\begin{eqnarray}  \label{transf}
U^{-1}(\vec a)\vec X_nU(\vec a)=\vec X_n+\vec a.
\end{eqnarray} 
For an infinitesimal translation the unitary operator can be written as $U(\vec a)=1+ i \vec a \cdot  \vec T + {\cal O}(\vec a \cdot \vec a)$ where $\vec T$ is an operator which generates the translation. This operator is called the momentum and we thus set $\vec T=\vec P/\hbar$. The constant $\hbar$ at this stage is introduce to make $\vec P \cdot \vec a/\hbar$ dimensionless and its value is only determined experimentally (like $c$ or $G_N$). It must carry the dimensions  $m^2 kg/s$. One thus finds
\begin{eqnarray}
U(\vec a)=1-\frac{i}{\hbar} \vec P \cdot \vec a+ {\cal O}(\vec a \cdot \vec a).
\end{eqnarray} 
The requirement (\ref{transf})  implies 
\begin{eqnarray}
\frac{i}{\hbar}[\vec P \cdot a, \vec X_n]=\vec a
\end{eqnarray} 
for any infinitesimal $\vec a$. One thus has
\begin{eqnarray} \label{comm}
[X_{ni},P_j] = i \hbar \delta_{ij},
\end{eqnarray}
where $X_i$ is the $i$-th component of $\vec X$ and $P_j$ is the $j-th$ component of $\vec P$ which is the total momentum of the system (if there are several particles in the problem).

We can now see that if we look at a one-particle state $\Phi(x)$, and define its position by 
\begin{eqnarray}
 X_i \Phi(x)=x_i \Phi(x)
\end{eqnarray}
the commutator (\ref{comm}) implies that 
\begin{eqnarray}
P_i \Phi(x)= i \hbar \frac{\partial}{\partial x_j} \Phi(x).
\end{eqnarray}

In principle we could have a different $\hbar$ for the different components of $P_j$. We need to assume rotational invariance which will lead to a universal $\hbar$. The requirement of time translation invariance leads to the concept of energy  and thus Hamiltonian and thus to the  Schr\"odinger equation:
\begin{eqnarray}
i \hbar \frac{\partial}{\partial t} \Psi = H \Psi 
\end{eqnarray}
note that at this stage the $\hbar$ of $P$ and that of $H$ could be different. It is only when one requests invariance under Galilean transformation that one finds that the two $\hbar$s must be the same.
Galilean invariance (and there is a similar one when Lorentz invariance is imposed), implies
\begin{eqnarray}
[\vec K, H]=-i \vec P
\end{eqnarray}
where $\vec K$ is the boost generator. This relation forces one to assume
\begin{eqnarray}
H=\sum_n \frac{\vec P_n \cdot \vec P_n}{2 m_n} + V
\end{eqnarray}
where $V$ is a function called the potential which only depends on the differences of the particle coordinate vectors. The constant $\hbar$ is thus universal in quantum mechanics.

A similar reasoning applies  to quantum field theory. Following the derivation of \cite{Weinberg:1995mt}, we consider one-particle states. They are defined by
\begin{eqnarray}
P^\mu \Psi_{p,\sigma} = p^\mu \Psi_{p,\sigma}
\end{eqnarray}
where $P^\mu$ is the four-momentum operator, $p^\mu$ is the corresponding eigenvalue  and $\sigma$ stands for any other quantum number. Under translations, the one-particle state transforms as 
\begin{eqnarray}
U(a)  \Psi_{p,\sigma}= \exp(-\frac{i}{\hbar_2} p \cdot a)  \Psi_{p,\sigma}
\end{eqnarray}
again $\hbar_2$ here is introduced to compensate the mass dimensions of $p \cdot a$.

The creation operator $a^\dagger$ is defined by the way it acts on a multiple particle state:
\begin{eqnarray}
a^\dagger (q) \Phi_{q_1 q_2 ... q_N} \equiv \Phi_{q q_1 q_2 ... q_N}
\end{eqnarray}
the creation operator adds a particle with quantum number $q$ to the multiple particle state.
The annihilation operator $a(q)$ is defined by
\begin{eqnarray}
a (q) \Phi_{q_1 q_2 ... q_N} = \sum_{r=1}^{N} (\pm)^{r+1} \delta(q-q_r) \Phi_{q_1...q_{r-1} q_{r+1}... q_N}
\end{eqnarray}
with a $+1$ sign for bosons or fermions respectively. One can easily show that 
\begin{eqnarray}
[ a (q_1) a^\dagger (q_2) \mp a^\dagger (q_1) a (q_2)] = \delta(q_1-q_2)  
\end{eqnarray}
where the top sign applies to bosons while the bottom one to fermions.
One can then define creation $\psi_l^-$ and annihilation fields $\psi_l^+$
\begin{eqnarray}
\psi_l^+ &=& \sum \int d^3p\  u_l(x;\vec p,\sigma, n) a(\vec p,\sigma, n) \\
\nonumber
\psi_l^- &=& \sum \int d^3p \ v_l(x;\vec p,\sigma, n) a^\dagger(\vec p,\sigma, n). 
\end{eqnarray}
The requirement of invariance under a translation leads to the requirement that
\begin{eqnarray}
u_l(x;\vec p,\sigma,n) &=&(2 \pi \hbar )^{-3/2} e^{\frac{i}{\hbar_2} p \cdot x} u_l(\vec p,\sigma,n) \\
\nonumber
v_l(x;\vec p,\sigma,n) &=&(2 \pi \hbar)^{-3/2} e^{-\frac{i}{\hbar_2} p \cdot x} v_l(\vec p,\sigma,n).
\end{eqnarray}
Fields are classified according to representation of $SO(3,1)$. The properties of Lorentz invariance as well as C, P, T transformations lead to relations between $u_l$ and $v_l$ which depend on the representation of the Wigner little group under consideration. In any case, these relations insure that the same $\hbar$ appears in both expressions. 

We will now specialize to the case of scalar fields, but the reasoning can be trivially extended to any other fields. For a causal scalar field one finds:
\begin{eqnarray}
\phi^+(x)=\int \frac{d^3p}{(2 \pi \hbar)^{3/2}} \  \frac{1}{\sqrt{2 p^0}} a(\vec p) e^{\frac{i}{\hbar_2} p \cdot x} 
\end{eqnarray}
and 
\begin{eqnarray}
\phi^-(x)=\int \frac{d^3p}{(2 \pi \hbar)^{3/2}} \  \frac{1}{\sqrt{2 p^0}} a^\dagger(\vec p) e^{-\frac{i}{\hbar_2} p \cdot x} =\phi^{+\dagger}(x).
\end{eqnarray}
The scalar field is then defined by
\begin{eqnarray}
\phi(x)= \int \frac{d^3p}{ (2\pi \hbar)^{3/2} (2 p^0)^{1/2}}\left [a(\vec p) e^{\frac{i}{\hbar_2} p \cdot x} + a^\dagger(\vec p) e^{-\frac{i}{\hbar_2} p \cdot x} \right ].
\end{eqnarray}
One finds
\begin{eqnarray}
[\phi(x),\phi^\dagger(y)]=\Delta(x-y)
\end{eqnarray}
with 
\begin{eqnarray}
\Delta(x)= \hbar_2 \int \frac{d^3k}{ (2\pi)^{3} 2 k^0} \left [e^{i k \cdot x} - e^{-i k \cdot x} \right].
\end{eqnarray}
One can show that 
\begin{eqnarray}
\Delta(\vec x,0)=0\ \  \mbox{and} \ \ \dot \Delta(\vec x,0)= - i \hbar_2 \delta^3(\vec x),
\end{eqnarray}
where the dot above a variable denotes the derivative with respect to $x^0$. One then verifies
\begin{eqnarray}
[\phi(\vec x,t),\dot \phi(\vec y,t)]= i \hbar_2 \delta^3(\vec x-\vec y).
\end{eqnarray}
One may now wonder whether $\hbar$ should be the same for all fields and the same in quantum mechanics and quantum field theory. The underlying reason for the universality of $\hbar$ is that it appears in the quantum Lorentz algebra

An infinitesimal Lorentz transformation is given by
\begin{eqnarray}
U(1+w,\epsilon)= 1 +\frac{i}{2 \hbar} w_{\mu\nu}  M^{\mu\nu} +\frac{i}{2 \hbar} \epsilon_\sigma P^\sigma.
\end{eqnarray}
One may worry that the $\hbar$s appearing in the transformation could be different, but one always has the freedom to normalize the transformations $w_{\mu\nu}$ and $\epsilon_\sigma$ in such a way that the same $\hbar$ appears in both terms of the equation. Let's calculate
\begin{eqnarray}
U(\Lambda,a) U(1+w,\epsilon)U^{-1}(\Lambda,a)=U(\Lambda(1+w)\Lambda^{-1},\Lambda \epsilon-\Lambda w\Lambda^{-1}a)
\end{eqnarray}
By comparing the coefficients of $w$ and $\epsilon$, one finds
\begin{eqnarray}
U(\Lambda,a)P^\mu  U^{-1}(\Lambda,a)=\Lambda_\nu^{\ \mu} P^\nu
\end{eqnarray}
\begin{eqnarray}
U(\Lambda,a) M^{\rho\sigma}  U^{-1}(\Lambda,a)=\Lambda_\mu^{\ \rho} \Lambda_\nu^{\ \sigma} (M^{\mu\nu}-a^\mu P^\nu +a^\nu P^\mu)
\end{eqnarray}
Let us now consider $\Lambda^\mu_{\ \nu} =\eta^\mu_{\ \nu} + w^\mu_{\ \nu}$, one finds
\begin{eqnarray}
i \left [\frac{1}{2\hbar} w_{\mu\nu} M^{\mu\nu} -\frac{1}{\hbar}\epsilon_\mu P^\mu, M^{\rho\sigma} \right ] = w_{\mu}^{\ \rho} M^{\mu\sigma} +w_{\nu}^{\ \sigma} M^{\rho\nu} -\epsilon^\rho P^\sigma+\epsilon^\sigma P^\rho
\end{eqnarray}
and
\begin{eqnarray}
i \left [\frac{1}{2\hbar} w_{\mu\nu} M^{\mu\nu} -\frac{1}{\hbar}\epsilon_\mu P^\mu, P^{\rho} \right ] = w_{\mu}^{\ \rho} P^{\mu}
\end{eqnarray}
which leads to the Lie Algebra of the Lorentz group
\begin{eqnarray}
[M^{\mu\nu},M^{\rho\sigma}] &=& i \hbar (g^{\mu\rho} M^{\nu\sigma}-g^{\nu\rho} M^{\mu\sigma} - g^{\mu\sigma} M^{\nu\rho}+g^{\nu\sigma} M^{\mu\rho}) \\ \nonumber
[P^{\mu},M^{\rho\sigma}] &=& i \hbar (g^{\mu\sigma} P^{\rho} -g^{\mu\rho} P^{\sigma})\\ \nonumber
[P^{\mu},P^{\nu}]&=&0
\end{eqnarray}
and we see that the $\hbar$ appearing in the algebra must be the same in the first two lines.

Let's now introduce the angular momentum operator $\vec J$ is given by $J_i=\epsilon_{ijk} M^{jk}$ and the boost operator is given by $K_i=M^{i0}$. 
\begin{eqnarray}
[J_i,J_j]&=&i \hbar \epsilon_{ijk} J_k \\ \nonumber
[J_i,K_j]&=&i \hbar \epsilon_{ijk} K_k\\ \nonumber
[K_i,K_k]&=&-i \hbar \epsilon_{ijk} J_k\\ \nonumber
[J_i,H]&=&0\\ \nonumber
[J_i,P_j]&=&i \hbar \epsilon_{ijk} P_k \\ \nonumber
[K_i,H]&=&i \hbar P_i\\ \nonumber
[K_i,P_j]&=&i \hbar \delta_{ijk} H\\ \nonumber
[P_i,P_j]&=&0\\ \nonumber
[P_i,H]&=&0.
\end{eqnarray}

Thus $\hbar$ must be universal to quantum mechanics and quantum field theory if we consider that quantum mechanics is the low velocity limit of relativistic quantum mechanics. Furthermore, if quantum gravity introduces an energy dependence of $\hbar$ this energy dependence must be universal and affect all processes.  Note that if Lorentz symmetry is violated at high energies, then it is conceivable for $\hbar$ to be non-universal at these energies. Let us now study the phenomenology of an energy dependent $\hbar$ assuming that Lorentz invariance is a valid symmetry of nature. 

\section{What do we know about the energy dependence of $\hbar$?}

In this section, we shall investigate modifications of the quantization rules which can be parametrized phenomenologically by an energy scale dependent Planck's constant. We thus adopt
\begin{eqnarray}
[x_i,p_j]=i \hbar(\mu) \delta_{ij}
\end{eqnarray}
in quantum mechanics and e.g. 
\begin{eqnarray}
[\phi({\bf x}),\Pi({\bf y})]=i \hbar(\mu) \delta({\bf x} - {\bf y})
\end{eqnarray}
for a scalar field with an obvious generalization for fields carrying a different spin. This parametrization enables us to consider both frameworks, that is, we can both consider cases where the theory is becoming more classical as short distances are probed, i.e. if $\hbar(\mu)$ decreases with increasing energy scale $\mu$ or more quantum in a sense that $\hbar(\mu)$ increases with increasing energy scale. We shall consider two experiments. The first one is the anomalous magnetic moment of the muon which is a low energy experiment but one of the most sensitive one done to date. The second one is a cosmological observation, namely the cosmic microwave background which probes inflation. While there is a strong model dependence in the second one, this is a very high energy process (the scale of inflation is close to the Planck scale) and if one assumes that the fluctuations in temperature in the cosmic microwave background are due to fluctuations of the inflaton, one can rule out models where $\hbar$ would tend to zero at energy scales close to the Planck scale.

\subsection{The anomalous magnetic moment of the muon}

The non-relativistic form of the Dirac equation describing a spinor in an electromagnetic field is given by
\begin{eqnarray}
i \hbar \frac{\partial \phi}{\partial t} = \left ( \frac{\vec p^2}{2 m} -\frac{e}{2 m c} (\vec L + 2 \vec S) \cdot \vec B \right) \phi
\end{eqnarray}
where $\vec L=\vec r \times \vec p$ is the angular momentum and $\vec S$ is the spin of the spinor with eigenvalues $\pm \hbar(\mu)/2$. We see immediately that an energy dependence of $\hbar$ would account for an energy dependence of magnetic moment $g$  of fermions which was famously prediction by Dirac to be equal to 2. It is know that quantum gravitational effects modify the value of the magnetic moment. The anomalous magnetic moment is given by 
\begin{eqnarray}
a(\mu)= \frac{g(\mu)-2}{2}.
\end{eqnarray}
The anomalous magnetic moment of the muon is one of the best measured quantities \cite{PDG}
\begin{eqnarray}
a_{exp} = 11 659 209.1(5.4)(3.3) \times 10^{-10}
\end{eqnarray}
and also most precisely calculated quantities in the standard model of particle physics
\begin{eqnarray}
a_{SM} = 11 659 180.3(1)(42)(26) \times 10^{-10}
\end{eqnarray}
The difference between experiment and theory
$\Delta a = a_{exp} - a_{SM} = 288(63)(49) \times 10^{-11}$ allows us to set a limit on the energy dependence of $\hbar$. The relevant energy scale which is probed by the anomalous magnetic moment of the muon is that of the muon mass (105.7 MeV/c$^2$).
\begin{eqnarray}
\Delta \hbar =\hbar - \hbar(m_{muon}) < 3 \times 10^{-9} \ \mbox{eV}\cdot \mbox{s}
\end{eqnarray}
At 105.7 MeV, the deviation between a $\hbar$ which is energy dependent and its standard value cannot be larger than $3 \times 10^{-9}$. On the other hand, the small deviation between the standard model prediction and the experimental measurement of the anomalous magnetic moment of the muon which  currently represents 3.6 $\sigma$ deviation, could be interpreted as the first sign that $\hbar$ differs from its standard value at 105.7 MeV by $3 \times 10^{-9}$ eV$\cdot$s.

\subsection{Inflation}
While the previous example gives a bound on the scale dependence of $\hbar$ at relatively low energy scale, inflation is a process that takes place at a very high energy scale close to the Planck scale. The issue is that we are now dealing with observations and not experiments. In particular, any bound on $\hbar$ coming from observations in the cosmic microwave background will be dependent on the specific model of inflation. However, the energy scale of inflation is known to be close to $10^{16}$ GeV. A common feature of inflationary model is the prediction of the existence of fluctuation of the temperature in the cosmic microwave background corresponding to quantum fluctuation of the inflaton. Assuming that inflation really took place, the observation of these fluctuations leads to the conclusion that at an energy scale of $10^{16}$ GeV, physics is still described by a quantum theory. This implies that at this energy scale, $\hbar \neq 0$ as one would expect from a model of emergent quantum mechanics. Let us now study this quantitatively.

The standard normalization implies that the perturbation at the present Hubble scale $\delta_H$ is given by \cite{Liddle:1993ch,Liddle:1999mq}:
\begin{eqnarray}
\delta_H^2=\frac{32}{75} V G_N^2 (\hbar(\mu) c)^2 \frac{1}{\epsilon} \to \delta_H \sim 2 \times 10^{-5}
\end{eqnarray}
unless the slow role parameter  $\epsilon$ is anomalously small. The energy scale which is probed at the start of inflation is thus around $10^{16}$ GeV. When evaluating the slow role parameters
\begin{eqnarray}
\epsilon = \frac{\hbar(\mu) c}{16 \pi G_N} \left (\frac{V'}{V} \right )^2
\end{eqnarray}
and
\begin{eqnarray}
\eta = \frac{\hbar(\mu) c}{8 \pi G_N} \left (\frac{V''}{V} \right )
\end{eqnarray}
we should thus evaluate them using $\hbar(10^{16}$ GeV). By measuring the slow parameters, we can thus probe the value of $\hbar$ close to the Planck mass for a given model of inflation since the slow role parameters depend on the potential $V$ and thus of the model. One can also consider the number of e-folding which is given by
\begin{eqnarray}
N= \frac{8 \pi G_N}{\hbar(\mu) c} \int^{\phi_N}_{\phi_e} d \phi \frac{V}{V'}.
\end{eqnarray}
It is in the range 50-60 and used to extract $\phi_N$. The value of the inflation field at the end of inflation $\phi_e$ calculated from the requirement that $\epsilon>1$ and $\eta>1$ and is also dependent on the sliding $\hbar$. In the limit of $\hbar \to 0$, the slow role parameters go to zero. Assuming that the temperature fluctuations in the cosmic microwave background are really due to the fluctuations of the inflaton, we can rule out from current observations that $\hbar(10^{16}$ GeV)$= 0$.

While small changes of $\hbar$ will be tough to observe and strongly model dependent, we can conclude that at $10^{16}$ GeV, nature is still described by quantum mechanics. If quantum mechanics is emergent, as it would be in 't Hooft's model, we must assume that the underlying theory becomes relevant at a very high energy scale above $10^{16}$ GeV.

\section{Conclusion}
In this paper we have investigated a possible energy scale dependence of the quantization rules and in particular, from a phenomenological point of view, an energy scale dependence of an effective $\hbar$. Different models of quantum mechanics and quantum gravity lead to an $\hbar$ which is effectively scale dependent. We have shown that this effective $\hbar$ must be universal since modern theories of the world are based on the Lorentz algebra. We have set a bound on the deviation from the value of $\hbar$ at the muon scale from its usual value using measurements of the anomalous magnetic moment of the muon.  Assuming that inflation took place, we can conclude that nature is described by a quantum theory at least up to an energy scale of about $10^{16}$ GeV. Finally it is worth mentioning that an energy scale dependence of Planck's constant has some interesting consequences for thought experiments probing an unification of general relativity and quantum mechanics. Firstly, it would lead to an energy scale dependence of the Planck mass which is given by $\sqrt{\hbar c/G_N}$. Note that Newton's constant fixes the strength of gravitational interactions, the limit $\hbar \to 0$, thus does not imply that the strength of gravitation goes to zero. Secondly, the Planck length $l_P=\sqrt{\hbar G/c^3}$, which can be shown to be the minimal length \cite{Calmet:2004mp} that can be measured given our current understanding of quantum mechanics and general relativity, goes to zero in the limit where $\hbar$ goes to zero. This can be easily understood since the standard quantum limit goes to zero in that limit. In the limit $\hbar \to 0$, the minimal length or Planck length tends to zero. This would be a problem for a quantum mechanical description of general relativity. It is often conjectured that singularities in general relativity are cured or at least hidden by the minimal length. If the minimal length tends to zero, singularities may be observable.

\section*{Acknowledgment}
This work is supported in part by the Science and Technology Facilities Council (grant number  ST/J000477/1). 
 

\bigskip{}

\end{document}